\documentclass[prl,twocolumn,superscriptaddress,floatfix,amsmath,footinbib,amssymb]{revtex4}

\usepackage{amssymb}

\usepackage{graphicx}
\usepackage{grffile}
\usepackage{amsmath}
\usepackage{amsbsy}
\usepackage{amsthm}
\usepackage{bbm}
\usepackage{bm}
\usepackage{epsfig}

\begin{document}

\title{Half-page derivation of the Thomas precession}
\date{\today}

\author{Andrzej Dragan}
\affiliation{Institute of Theoretical Physics, University of Warsaw, Ho\.{z}a 69, 00-049 Warsaw, Poland}

\author{Tomasz Odrzyg\'{o}\'{z}d\'{z}}
\affiliation{College of Inter-Faculty Individual Studies in Mathematics and Natural Sciences, University of Warsaw, \.{Z}wirki i Wigury 93, Warsaw, Poland}

\maketitle

\emph{Introduction.\thinspace---} Composition of two non-collinear Lorentz boosts, results in a Lorentz transformation that is not a pure boost but a composition of a boost and a spatial rotation, known as the Wigner rotation \cite{Wigner}. As a consequence, a body moving on a curvilinear trajectory undergoes a rotational precession, that was first discovered by Thomas \cite{Thomas}. In the vast majority of textbooks this phenomenon is either omitted or described with very sophisticated mathematical tools, such as gyrogroups, associative-commutative groupoids, etc. \cite{a}. Here we present a half-page derivation of the Thomas precession formula using only basic vector operations. Our approach is not only simple and clear, but also builds a better physical intuition of this relativistic effect.

\emph{Derivation.\thinspace---} 
Let us introduce three inertial observers: Alice, Bob and a cat and denote their reference frames by $A$, $B$, and $C$, respectively. We choose them such that $A$ is non-rotated with respect to $B$, and $B$ is non-rotated with respect to $C$ (however in general $C$ is inevitably going to be rotated with respect to $A$). Let Bob hold the cat and move with a constant velocity $\boldsymbol{v}$ with respect to Alice. Unfortunatelly, at some point the cat decides to run away from Bob with an infinitesimally small velocity $\mbox{d}\boldsymbol{v'}$ with respect to him. It follows that Bob is moving with velocity $-\mbox{d}\boldsymbol{v'}$ with respect to the cat and Alice is moving with velocity $-\boldsymbol{v}$ with respect to Bob - see Fig.~\ref{ABC}. Let us denote the cat's velocity in Alice's frame by $\boldsymbol{v}+\mbox{d}\boldsymbol{v}$. Since the cat is rotated relative to Alice, her velocity in the cat's frame $\widetilde{\boldsymbol{v}}$ is yet to be determined: $\widetilde{\boldsymbol{v}}\neq -\boldsymbol{v}-\mbox{d}\boldsymbol{v}$. The angle $\mbox{d}\boldsymbol{\Omega}$ of that rotation equals:
\begin{equation}
\label{domega}
\mbox{d}\boldsymbol{\Omega} =-\frac{\widetilde{\boldsymbol{v}}}{|\widetilde{\boldsymbol{v}}|} \times \frac{\boldsymbol{v}+\mbox{d}\boldsymbol{v}}{|\boldsymbol{v}+\mbox{d}\boldsymbol{v}|} \approx -\frac{1}{v^2} \widetilde{\boldsymbol{v}} \times (\boldsymbol{v}+\mbox{d}\boldsymbol{v}).
\end{equation}
To derive the formula for the precession rate in Alice's frame we use the velocity composition law, which for the simplest case of motion along the $x$ axis with the velocity $V$ is given by: 
\begin{equation}
\label{SzczegolnaTransf}
u'^x = \frac{u^x-V}{1-\frac{u^xV}{c^2}},~~~
u'^y = \frac{u^y\sqrt{1-\frac{V^2}{c^2}}}{1-\frac{u^xV}{c^2}},~~~
u'^z = \frac{u^z\sqrt{1-\frac{V^2}{c^2}}}{1-\frac{u^xV}{c^2}}, \nonumber
\end{equation}
where $\boldsymbol{u}$ and $\boldsymbol{u'}$ are velocities of some object observed from the rest frame and a moving frame, respectively. For an arbitrary velocity $\boldsymbol{V}$ of the moving frame the transformation takes the form:
\begin{equation}
\label{OgolnaTransf} \boldsymbol{u'} = \frac{\sqrt{1-\frac{V^2}{c^2}}\left(\boldsymbol{u}-\frac{\boldsymbol{u}\cdot\boldsymbol{V} }{V^2}\boldsymbol{V}\right)-\left(\boldsymbol{V}-\frac{\boldsymbol{u}\cdot\boldsymbol{V} }{V^2}\boldsymbol{V}\right) }{1-\frac{\boldsymbol{u}\cdot\boldsymbol{V}}{c^2}}.
\end{equation}
where it is assumed that the primed and unprimed frames are mutually non-rotated. We will now follow two simple steps in order to express $\tilde{\boldsymbol{v}}$ appearing in Eq.~\eqref{domega} in terms of $\boldsymbol{v}$ and $\mbox{d}\boldsymbol{v}$. First we describe how the observers $A$ and $B$ observe $C$, and second how the observers $B$ and $C$ observe $A$. In the first step we use Eq.~\eqref{OgolnaTransf} to describe the transition from the frame $A$ to $B$, which involves the following substitutions in the formula \eqref{OgolnaTransf}: $\boldsymbol{V}\rightarrow\boldsymbol{v}$, $\boldsymbol{u}\rightarrow\boldsymbol{v}+\mbox{d}\boldsymbol{v}$, and $\boldsymbol{u'}\rightarrow\mbox{d}\boldsymbol{v'}$. After leaving only the first order terms in $\mbox{d}\boldsymbol{v}$ we get: 
\begin{equation}
\label{dvprim}
\mbox{d}\boldsymbol{v'} \approx \frac{1}{\sqrt{1-\frac{v^2}{c^2}}}\left(\mbox{d}\boldsymbol{v}-\frac{\boldsymbol{v}\cdot\mbox{d}\boldsymbol{v}}{v^2}\boldsymbol{v}\right)+\frac{1}{1-\frac{v^2}{c^2}}\frac{\boldsymbol{v}\cdot\mbox{d}\boldsymbol{v}}{v^2}\boldsymbol{v}.
\end{equation}
Secondly, we use again use the Eq.~\eqref{OgolnaTransf} applied to the transition from $B$ to $C$ that involves the following substitutions in \eqref{OgolnaTransf}: $\boldsymbol{V}\rightarrow\mbox{d}\boldsymbol{v'}$, $\boldsymbol{u}\rightarrow-\boldsymbol{v}$, and $\boldsymbol{u'}\rightarrow\widetilde{\boldsymbol{v}}$. Droping out higher order terms in $\mbox{d}\boldsymbol{v'}$ leads to: 
\begin{figure}
\begin{center}
\epsfig{file=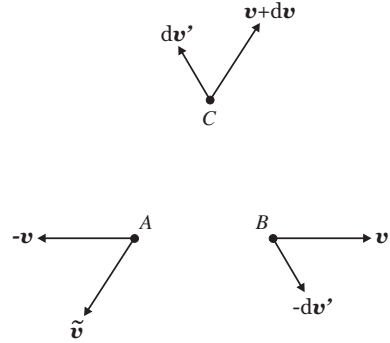, width=5cm} \caption{\label{ABC} Schematic diagram of mutual velocities between Alice, Bob and the cat.}
\end{center}
\end{figure}
\begin{equation}
\label{vtilde}
\widetilde{\boldsymbol{v}} \approx -\boldsymbol{v} + \frac{\boldsymbol{v}\cdot\mbox{d}\boldsymbol{v'}}{c^2}\boldsymbol{v} - \mbox{d}\boldsymbol{v'}.
\end{equation}
Substituting \eqref{dvprim} to \eqref{vtilde}, then everything to \eqref{domega} and dividing both sides of the resulting equation by $\mbox{d}t$ we obtain: 
\begin{equation}
\dot{\boldsymbol{\Omega}} = -\frac{1}{v^2}\left(\frac{1}{\sqrt{1-\frac{v^2}{c^2}}}-1\right) \boldsymbol{v}\times\dot{\boldsymbol{v}}.
\end{equation}
The above formula expresses the well-known Thomas precession of a body moving with variable velocity $\boldsymbol{v}(t)$ and observed by a fixed, inertial observer.

\end{document}